\documentclass{ws-ijmpa}

\begin{document}

\newcommand{\tr}[0]{{\rm tr}}
\def\rescale{\fontsize{6}{1}}

\markboth{M.F.M. Lutz and J. Hofmann} {Dynamically generated
hidden-charm baryon resonances}

%
\catchline{}{}{}{}{}
%

\title{Dynamically generated hidden-charm baryon resonances}

\author{M.F.M.Lutz and J. Hofmann}

\address{Gesellschaft f\"ur Schwerionenforschung (GSI), Planck Str. 1, 64291 Darmstadt, Germany, m.lutz@gsi.de}

\maketitle

\begin{history}
\received{Day Month Year}
\revised{Day Month Year}
\end{history}

\begin{abstract}
Identifying a zero-range exchange of vector mesons as the driving
force for the s-wave scattering of pseudo-scalar mesons off the
baryon ground states, a rich spectrum of hadronic nuclei is
formed. We argue that chiral symmetry and large-$N_c$
considerations determine that part of the interaction which
generates the spectrum. We suggest the existence of strongly bound
crypto-exotic baryons, which contain a charm-anti-charm pair. Such
states are narrow since they can decay only via OZI-violating
processes.

\end{abstract}

\ccode{PACS numbers: 12.38 Cy, 12.38 Lg, 12.39 Fe}

\section{Introduction}

\begin{figure}[t]
\epsfxsize=13.5cm   
\epsfbox{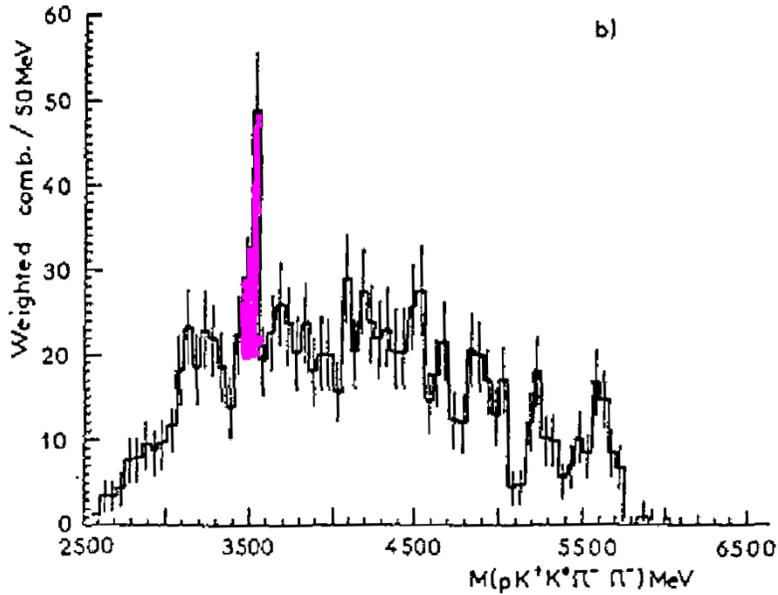}
\caption{Measured five-body invariant mass distribution in a 19
GeV $\pi^-$ beam experiment at CERN. The figure is taken from Ref.$\,6$}
\label{fig:Nstar}
\end{figure}

The existence of strongly bound crypto-exotic baryon systems with
hidden charm would be a striking feature of strong interactions
\cite{Brodsky:Schmidt:Teramond:90,Kaidalov:Volkovitsky:92,Gobbi:Riska:Scoccola:92}.
Such states may be narrow since their strong decays are
OZI-suppressed \cite{Landsberg:94}. Indeed, a high statistics
bubble chamber experiment performed 30 years ago with a $K^-$ beam
reported on a possible signal for a hyperon resonance of mass 3.17
GeV of width smaller than 20 MeV \cite{Amirzadeh:79}.  About ten
years later a further bubble chamber experiment using a high
energy $\pi^-$ beam suggested a nucleon resonance of mass 3.52 GeV
with a narrow width of $7^{+20}_{-7}$ MeV . In Fig.
\ref{fig:Nstar} we recall the measured five body
($p\,K^+\,K^0\,\pi^-\pi^-$) invariant mass
distribution\cite{Karnaukhov:91} suggesting the existence of a
crypto-exotic nucleon resonance.

It is the purpose of the present talk to review a study addressing
the possible existence of crypto-exotic baryon
systems\cite{Hofmann:Lutz:2005,Hofmann:Lutz:2006}. In view of the
highly speculative nature of such states it is important to
correlate the properties of such states to those firmly
established, applying a unified and quantitative framework. We
extended previous works Ref. \refcite{LK04-charm,LK05} that
performed a coupled-channel study of the s-wave scattering
processes where a Goldstone boson hits a baryon ground state. The
spectrum of $J^P=\frac{1}{2}^-$ and $J^P=\frac{3}{2}^-$ molecules
obtained in Ref. \refcite{Granada,Copenhagen,LK04-charm,LK05} is
quite compatible with the so far observed states. Analogous
computations successfully describe the spectrum of zero and
open-charm mesons with $J^P=0^+$ and $1^+$ quantum numbers
\cite{LK04-axial,KL04,HL04}. These developments were driven by the
hadrogenesis conjecture: meson and baryon resonances that do not
belong to the large-$N_c$ ground state of QCD should be viewed as
hadronic nuclei \cite{LK02,LWF02,Granada,Copenhagen,LK04-axial}.
Generalizing those computations to include D- and $\eta_c$-mesons
in the intermediate states offers the possibility to address the
formation of crypto-exotic baryon states (see also Ref.
\refcite{Mo:80,Rho:Riska:Scoccola:92,Min:Oh:Rho:95,Oh:Kim:04}).

The results of Ref. \refcite{Granada,Copenhagen,LK04-charm,LK05}
were based on the leading order chiral Lagrangian, that predicts
unambiguously the s-wave interaction strength of Goldstone bosons
with  baryon ground states in terms of the pion decay constant.
Including the light vector mesons as explicit degrees of freedom
in a chiral Lagrangian gives an interpretation of the leading
order interaction in terms of the zero-range t-channel exchange of
light vector mesons \cite{Weinberg:68,Wyld,Dalitz,sw88,Bando:85}.
The latter couple universally to any matter field in this type of
approach. Based on the assumption that the interaction strength of
D- and $\eta_c$-mesons with the baryon ground states is also
dominated by the t-channel exchange of the light vector mesons, we
performed a coupled-channel study of crypto-exotic baryon
resonances\cite{Hofmann:Lutz:2005,Hofmann:Lutz:2006}.

\section{S-wave baryon resonances with zero charm}

The spectrum of $J^P=\frac{1}{2}^- $ baryon resonances as
generated by the t-channel vector-meson exchange interaction via
coupled-channel dynamics falls into two types of states.
Resonances with masses above 3 GeV couple strongly to mesons with
non-zero charm content. In the SU(3) limit those states form an
octet and a singlet. All other states have masses below 2 GeV. In
the SU(3) limit they group into two degenerate octets and one
singlet. The presence of the heavy channels does not affect that
part of the spectrum at all (see Ref.
\refcite{Hofmann:Lutz:2005}). We reproduce the success of previous
coupled-channel computations of Ref. \refcite{Granada,Copenhagen},
which predicts the existence of the s-wave resonances $N(1535),
\Lambda(1405), \Lambda(1670), \Xi(1690)$ unambiguously with masses
and branching ratios quite compatible with empirical information.
These states are omitted from Table 1.

Most spectacular are the resonances with hidden charm generated
above 3 GeV. The multiplet structure of such states is readily
understood. The mesons with $C=-1$  form a triplet which is
scattered off the $C=+1$ baryons being members of an anti-triplet
or sextet. We decompose the products into irreducible tensors
\begin{eqnarray}
3 \otimes \overline 3 = 1 \oplus 8 \,, \qquad
3 \otimes 6 = 8\oplus 10\,.
\label{3times6}
\end{eqnarray}
The coupled-channel interaction is attractive in the singlet for
the anti-triplet of baryons. Attraction in the octet sector is
provided by the sextet of baryons. The resulting octet of states
mixes with the $\eta'\,(N,\Lambda, \Sigma, \Xi)$ and
$\eta_c\,(N,\Lambda, \Sigma, \Xi)$ systems. A complicated mixing
pattern arises. All together the binding energies of the
crypto-exotic states are large. This is in part due to the large
masses of the coupled-channel states: the kinetic energy the
attractive t-channel force has to overcome is reduced.

\begin{table}[t]
\begin{center}
\tbl{ Spectrum of $J^P=\frac{1}{2}^-$ baryons with charm zero. The
3rd and 4th columns follow with SU(4) symmetric 3-point vertices.
In the 5th and 6th columns SU(4) breaking is introduced as
explained in the text. Only states with hidden charm are shown,
for a complete list see Ref. 7.} {\rescale
\setlength{\tabcolsep}{1.0mm} \setlength{\arraycolsep}{2.mm}
\renewcommand{\arraystretch}{0.75}
\begin{tabular}{|ll|c|c|c|c|c|}
\hline $C=0:$ &$ (\,I,\phantom{+}S)$  &
$\rm state$ &  $\begin{array}{c} M_R [\rm MeV]  \\ \Gamma_R \,[\rm MeV]  \end{array}$ &
$|g_R|$  & $\begin{array}{c} M_R [\rm MeV]  \\ \Gamma_R \,[\rm MeV]  \end{array}$ & $|g_R|$  \\
\hline
\hline
&$(\frac12,\phantom{+}0)$   &
$\begin{array}{l}  \pi\, N \\ \eta\, N \\ K \,\Lambda \\ K \,\Sigma \\
\eta' N \\ \eta_c N \\ \bar{D}\, \Lambda_c \\ \bar{D}\, \Sigma_c \end{array}$
& $\begin{array}{c} 3327 \\ 156 \end{array}$ & $\begin{array}{c} 0.1  \\ 0.1  \\ 0.1  \\ 0.1  \\ 1.4  \\ 0.7 \\ 0.5  \\ 5.7 \end{array}$
& $\begin{array}{c} 3520 \\ 7.3 \end{array}$ & $\begin{array}{c} 0.07 \\ 0.11 \\ 0.08 \\ 0.08 \\ 0.22 \\ 1.0 \\ 0.05 \\ 5.3 \end{array}$\\
\hline
&$(0,-1)$    &
$\begin{array}{l} \pi \,\Sigma \\ \bar{K}\, N \\ \eta\, \Lambda \\ K \,\Xi \\ \eta' \Lambda \\
\eta_c \Lambda \\ \bar{D}_s \Lambda_c \\ \bar{D}\, \Xi_c \\ \bar{D}\, \Xi_c'\end{array}$
& $\begin{array}{c} 3148 \\ 1.0  \end{array}$ & $\begin{array}{c} 0.04 \\ 0.03 \\ 0.03 \\ 0.04 \\ 0.08 \\ 0.08 \\ 3.2 \\ 5.0 \\ 0.1  \end{array}$
& $\begin{array}{c} 3234 \\ 0.57 \end{array}$ & $\begin{array}{c} 0.04 \\ 0.03 \\ 0.03 \\ 0.04 \\ 0.01 \\ 0.06 \\ 3.0 \\ 5.0 \\ 0.01 \end{array}$\\
\cline{3-7}
&$(0,-1)$    &
$\begin{array}{l} \pi\, \Sigma \\ \bar{K}\, N \\ \eta \,\Lambda \\ K \,\Xi \\
\eta' \Lambda \\ \eta_c \Lambda \\ \bar{D}_s \Lambda_c \\ \bar{D} \,\Xi_c \\ \bar{D}\, \Xi_c'\end{array}$
& $\begin{array}{c} 3432 \\ 161 \end{array}$ & $\begin{array}{c} 0.1  \\ 0.0  \\ 0.0  \\ 0.1  \\ 1.3  \\ 0.7  \\ 0.6  \\ 0.1  \\ 5.6 \end{array}$
& $\begin{array}{c} 3581 \\ 4.9 \end{array}$ & $\begin{array}{c} 0.06 \\ 0.01 \\ 0.03 \\ 0.07 \\ 0.20 \\ 0.93 \\ 0.05 \\ 0.02 \\ 5.3 \end{array}$\\
\hline
&$(1,-1)$    &
$\begin{array}{l} \pi \,\Lambda \\ \pi\, \Sigma \\ \bar{K}\, N \\ \eta\, \Sigma \\ K \,\Xi \\
\eta' \Sigma \\ \eta_c \Sigma \\ \bar{D}\, \Xi_c \\ \bar{D}_s \Sigma_c \\ \bar{D}\, \Xi_c' \end{array} $
& $\begin{array}{c} 3602 \\ 227 \end{array}$ & $\begin{array}{c} 0.1   \\ 0.1  \\ 0.2  \\ 0.1  \\ 0.1  \\ 1.5  \\ 1.2 \\ 0.6  \\ 4.6 \\ 2.9 \end{array}$
& $\begin{array}{c} 3930 \\ 11  \end{array}$ & $\begin{array}{c} 0.08  \\ 0.04 \\ 0.12 \\ 0.08 \\ 0.06 \\ 0.27 \\ 1.8 \\ 0.11 \\ 3.6 \\ 2.4 \end{array}$ \\
\hline
&$(\frac12,-2)$  &
$\begin{array}{l} \pi\, \Xi \\ \bar{K}\, \Lambda \\ \bar{K}\, \Sigma \\ \eta\, \Xi \\ \eta' \Xi \\
\eta_c \Xi \\ \bar{D}_s \Xi_c \\ \bar{D}_s \Xi_c' \\ \bar{D}\, \Omega_c \end{array}$
& $\begin{array}{c} 3624 \\ 204 \end{array}$ & $\begin{array}{c} 0.1  \\ 0.1  \\ 0.1  \\ 0.0  \\ 1.4  \\ 1.0 \\ 0.6  \\ 3.3 \\ 4.3 \end{array}$
& $\begin{array}{c} 3798 \\ 6.0 \end{array}$ & $\begin{array}{c} 0.08 \\ 0.04 \\ 0.04 \\ 0.01 \\ 0.22 \\ 1.2 \\ 0.10 \\ 2.9 \\ 4.0 \end{array}$ \\
\hline
\end{tabular}}
\label{tab:charm0a}
\end{center}
\end{table}

The states are narrow as a result of the OZI rule. The mechanism
is analogous to the one explaining the long life time of the
$J/\Psi$-meson. We should mention, however, a caveat. It turns out
that the width of the crypto-exotic states is quite sensitive to
the presence of channels involving the $\eta'$ meson. This is a
natural result since the $\eta'$ meson is closely related to the
$U_A(1)$ anomaly giving it large gluonic components. The latter
works against the OZI rule. We emphasize that switching off the
t-channel exchange of charm or using the SU(4) estimate for the
latter, strongly bound crypto-exotic states are formed. In Table
\ref{tab:charm0a} the zero-charm spectrum insisting on SU(4)
estimates for the couplings is shown in the 3rd and 4th column.
The mass of the crypto-exotic nucleon resonance comes at 3.33 GeV
in this case. Its width of 160 MeV is completely dominated by the
$\eta' N$ decay. The properties of that state can be adjusted
easily to be consistent with the empirical values claimed in Ref.
\refcite{Karnaukhov:91}. The $\eta'$ coupling strength to the
open-charm mesons can be turned off by an appropriate SU(4)
breaking. As a result the width of the resonance is down to about
1-2 MeV. It is stressed that the masses of the crypto-exotic
states are not affected at all. The latter are increased most
efficiently by allowing an OZI violating $\phi_\mu \,D \bar D $
vertex. All together it is possible to tune the mass and width of
the crypto-exotic nucleon at $3.52$ GeV and $7$ MeV. The result of
this scenario is shown in the last two rows of Table
\ref{tab:charm0a}. Further crypto-exotic states, members of the
aforementioned octet, are predicted at mass 3.58 GeV $(0,-1)$ and
3.93 GeV $(1,-1)$. The multiplet is completed with a
$(\frac{1}{2},-2)$ state at 3.80 GeV. The decay widths of these
states center around $\sim 7$ MeV. This  reflects the dominance of
their decays into channels involving the $\eta'$ meson. The
coupling constants to the various channels are included in Table
\ref{tab:charm0a}. They confirm the interpretation that the
crypto-exotic states discussed above are a consequence of a
strongly attractive force between the charmed mesons and the
baryon sextet.

A crypto-exotic SU(3) singlet state is formed due to strong
attraction in the $(\bar D_s\Lambda_c), (\bar D\,\Xi_c )$ system.
Its nature is quite different as compared to the one of the octet
states. This is because its coupling to the $\eta' \Lambda$
channel is largely suppressed.  We identify this state with a
signal claimed in the $K^-p$ reaction, where a narrow hyperon
state with 3.17 GeV mass and width smaller than 20 MeV was seen
\cite{Amirzadeh:79}. Using values for the coupling constants as
suggested by SU(4) the state has a mass and width of 3.148 GeV and
1 MeV (see 3rd and 4th column of Table \ref{tab:charm0a}).

\section{$J^P=\frac{3}{2}^-$ resonances with zero charm}

We turn to the resonances with $J^P=\frac32^-$. The spectrum falls
into two types of states. Resonances with masses above 3 GeV
couple strongly to mesons with non-zero charm content. In the
SU(3) limit those states form an octet. All other states have
masses below 2.5 GeV. In the SU(3) limit they group into an octet
and decuplet. The presence of the heavy channels does not affect
that part of the spectrum at all. We reproduce the previous
coupled-channel computation \cite{Copenhagen}.

Again an octet of resonances with hidden charm above 3 GeV is
formed. The multiplet structure of such states is readily
understood. The mesons with $C=-1$  form a triplet which is
scattered off the charmed baryons forming a sextet, the
decomposition into irreducible tensors is the same as in
(\ref{3times6}). The interaction is attractive in the
crypto-exotic octet and repulsive in the decuplet. For the
formation of the states the charm-exchange processes are
irrelevant. This holds as long as the SU(4) estimates for the
coupling constants are correct within a factor of three. Thus the
crypto-exotic sector may be characterized by the decomposition
\begin{eqnarray}
\frac{1}{4\,g^2}\,\sum_{V\in [9]}\,C^{(C=0)}_V =
C^{\rm crypto}_{[8]} - 2\, C^{\rm crypto}_{[10]}\,,
\label{crypto-decomposition}
\end{eqnarray}
where we assume chiral and large-$N_c$ relations for the
three-point vertices again. The binding energies of the
crypto-exotic states are large.

In Tab. \ref{tab:charm0b} the spectrum of crypto-exotic baryons is
shown. The charm-exchange contributions are estimated by a SU(4)
ansatz for the 3-point vertices. The octet of states is narrow as
a result of the OZI rule. The precise values of the width
parameters are sensitive to the SU(4) estimates. In contrast to
the $J^P=\frac12^-$ states, channels that involve the $\eta'$
meson do not to play a special role in the d-wave spectrum. This
is in part a consequence that the $\eta'$ channels decouple from
the crypto-exotic sector in the SU(3) limit. It is interesting to
observe that a narrow nucleon resonance  is predicted at 3.42 GeV.
One may speculate that the crypto-exotic resonance claimed at 3.52
GeV in \cite{Karnaukhov:91} may be a d-wave rather than s-wave
state. If so its decay into the $\eta' N$ channel should be
suppressed. Given the uncertainties of the claim
\cite{Karnaukhov:91} we refrain from fine tuning the model
parameters as to push up the $(\frac{1}{2},0)$ state.

The results collected in Tab. \ref{tab:charm0a},\ref{tab:charm0b} are
subject to large uncertainties. The evaluation of the total
width as well as a more reliable estimate of the binding energies
of the crypto-exotic states requires the consideration of further
partial-wave contributions.  The large binding energy
obtained suggest a more detailed study that is based on
a more realistic interaction taking into account in particular the
finite masses of the t-channel exchange processes.

\begin{table}[t]
\begin{center}
\tbl{ Spectrum of $J^P=\frac{3}{2}^-$ baryons with charm zero.
Here all columns follow with SU(4) symmetric 3-point vertices.
Only states with hidden charm are shown, for a complete list see
Ref. 8.} {\rescale \setlength{\tabcolsep}{1.0mm}
\setlength{\arraycolsep}{2.mm}
\renewcommand{\arraystretch}{0.75}
\begin{tabular}{|ll|c|c|c|}
\hline $C=0:$ &$ (\,I,\phantom{+}S)$  & $\rm state$ &
$\begin{array}{c} M_R [\rm MeV]  \\ \Gamma_R \,[\rm MeV] \end{array}$ & $|g_R|$  \\
\hline \hline &$(\frac12,\phantom{+}0)$   &
$\begin{array}{l} \pi\, \Delta \\ K\,\Sigma \\ \bar D\,\Sigma_c \end{array}$
& $\begin{array}{c} 3430 \\ 0.50 \end{array}$ & $\begin{array}{c} 0.05  \\ 0.04  \\ 5.6 \end{array}$ \\
\hline
&$(0,-1)$    &
$\begin{array}{l} \pi\, \Sigma \\ K \,\Xi \\ \bar D\,\Xi_c \end{array}$
& $\begin{array}{c} 3538 \\ 0.63 \end{array}$ & $\begin{array}{c} 0.04 \\ 0.05 \\ 5.5 \end{array}$ \\
\hline
&$(1,-1)$    &
$\begin{array}{l} \pi\,\Sigma \\ \bar K \,\Delta \\ \eta \,\Sigma \\ K\, \Xi \\\eta' \Sigma \\ \eta_c \Sigma \\ \bar{D}_s \Sigma_c \\ \bar{D}\, \Xi_c \end{array}$
& $\begin{array}{c} 3720 \\ 0.83 \end{array}$
& $\begin{array}{c} 0.02 \\ 0.04 \\ 0.04 \\ 0.03 \\ 0.01 \\ 0.20 \\ 4.5 \\ 2.8 \end{array}$ \\
\hline
&$(\frac12,-2)$    &
$\begin{array}{l} \pi\, \Xi \\ \bar K\, \Sigma \\ \eta\, \Xi \\ K\, \Omega \\ \eta' \Xi \\ \eta_c \Xi \\ \bar D_s \Xi_c \\ \bar D\, \Omega_c \end{array}$
& $\begin{array}{c} 3742 \\ 1.1 \end{array}$
& $\begin{array}{c} 0.02 \\ 0.03 \\ 0.04 \\ 0.06 \\ 0.03 \\ 0.16 \\ 3.2 \\ 4.2  \end{array}$ \\
\hline
\end{tabular}}
\label{tab:charm0b}
\end{center}
\end{table}

\section{Summary}

We reviewed coupled-channel studies of crypto-exotic s-wave and
d-wave baryon resonances with charm $0$. The dominant interaction
is defined by the exchange of light vector mesons in the
t-channel. All relevant coupling constants are obtained from
chiral and large-$N_c$ properties of QCD. Less relevant vertices
related to the t-channel forces induced by the exchange of charmed
vector mesons  were estimated by applying SU(4) symmetry. Most
spectacular is the prediction of narrow  baryons with charm zero
forming below 4 GeV. Such states contain a $c \bar c$ pair. Their
widths parameters are small due to the OZI rule, like it is the
case for the $J/\Psi$ meson. We predict octets of crypto-exotic
s-wave and d-wave  states. The s-wave resonances decay dominantly
into channels involving an $\eta'$ meson. An even stronger bound
crypto-exotic SU(3) singlet s-wave state is predicted to have a
decay width of about 1 MeV only.

\end{document}